\documentstyle[aps,prd]{revtex} 
\begin{document}

\def\SLC{{\Omega h^2}_{\rm SLC}}
\def\SQC{{\Omega h^2}_{\rm SQC}}
\def\NCC{{\Omega h^2}_{\rm NCC}}
\def\ACC{{\Omega h^2}_{\rm ACC}}
\def\IGC{{\Omega h^2}_{\rm IGC}}
\def\COA{{\Omega h^2}_{\rm COA}} 

\def\mchi{m_{\chi} }
\def\gev{\mbox{ GeV}} 
\def\tev{\mbox{ TeV}}
\def\dtgev{\,\mbox{GeV}^{-2}}
\def\dgev{{1\over \mbox{m}^2\cdot \mbox{yr}}}
\def\hgev{\,\mbox{m}^{-2}\cdot \mbox{yr}^{-1}} 
\def\etal{{\it {et\ al.}}} 
\title{Squark-, Slepton- and Neutralino-Chargino coannihilation 
	effects in the low-energy effective MSSM}

\author{V.A.~Bednyakov}
\address{Laboratory of Nuclear Problems,
         Joint Institute for Nuclear Research, \\ 
         141980 Dubna, Russia; E-mail: bedny@nusun.jinr.ru}

\author{H.V.~Klapdor-Kleingrothaus and V.~Gronewold}

\address{Max-Planck-Institut f\"{u}r Kernphysik, \protect\\
        Postfach 103980, D-69029, Heidelberg, Germany \protect\\[3mm]}

\maketitle 

\begin{abstract} 
	We calculate the neutralino relic density
	within the low-energy effective 
	Minimal Supersymmetric extension of the Standard Model (effMSSM) 
 	taking into account slepton-neutralino,
	squark-neutralino and neutralino/chargino-neutralino 
	coannihilation channels. 
	By including squark (stop and sbottom) coannihilation channels 
	we extend our comparative study to all allowed
	coannihilations and obtain the  general result 
	that all of them give 
	sizable contributions to the reduction of the 
	neutralino relic density.
	Due to these coannihilation processes 
	some models (mostly with large neutralino masses) 
	fall within into the cosmologically interesting region 
	for relic density, but other models 
	drop out at this region. 
	Nevertheless, the predictions for 
	direct and indirect dark matter detection rates 
	are not strongly affected by
	these coannihilation channels in the effMSSM. 
\end{abstract} 

\section{Introduction} 
	A variety of data ranging from galactic rotation curves to 
	large scale structure formation and the cosmic microwave background
	radiation imply a significant density 
	$0.1< \Omega h^2 <0.3$  
\cite{acc} of so-called cold dark matter (CDM). 
	Here $\Omega=\rho/\rho_c$ and 
	$\rho_c= 3 H^2/8\pi G_N$ is the critical closure density of 
	the universe, $G_N$ is the Newton constant and 
	$h$ is the Hubble constant in units of 100 km/s/Mpc. 
	It is generally believed that most of the CDM is made of 
	weakly-interacting massive particles (WIMPs)
\cite{kt90}.  
	A commonly considered candidate for the WIMP 
	is the lightest neutralino, provided  
	it is the lightest supersymmetric particle (LSP) 
\cite{jkg}
	in the Minimal Supersymmetric extension of the Standard Model (MSSM).  
	Four neutralinos in the MSSM, being mass eigenstates, 
	are mixtures of the 
	bino $\widetilde B$, 
	wino $\widetilde W$ and 
	higgsinos $\widetilde H_d^0$, $\widetilde H_u^0$,
	and the  LSP can be written as a composition
$\chi \equiv 
  \widetilde \chi_1 = N_{11} \widetilde B     + N_{12} \widetilde W 
                    + N_{13} \widetilde H_d^0 + N_{14} \widetilde H_u^0 
$,\
	where $N_{ij}$ are the entries of the neutralino mixing matrix. 
	In SUSY phenomenology one usually classifies neutralinos as 
	gaugino-like (with ${\cal P} \approx 1$),
	higgsino-like (with ${\cal P} \approx 0$), and mixed, 
	where the gaugino fraction is defined as 
	${\cal P} = |N_{11}|^2 + |N_{12}|^2$.

	In most approaches the LSP is stable due to R-parity conservation
\cite{susyreview}. 
	The neutralino, being massive, neutral and stable, 
	often provides a sizeable contribution to the relic density. 
	The contribution of neutralinos to the relic density is
	strongly  model dependent and varies by several orders of magnitude 
	over the whole allowed parameter space of the MSSM.
	The neutralino 
	relic density then can impose stringent constraints on the
	parameters of the MSSM and the SUSY particle spectrum,
	and may have 
	important consequences both for studies of SUSY 
	at colliders and in astroparticle experiments.
	In the light of this and taking into account the 
	continuing improvements in determining the abundance of CDM
	and other components of the Universe, which have now reached an
	unprecedented precision 
\cite{cmb},  one needs 
	to perform an accurate enough
	computation of the WIMP relic abundance, which would allow 
	reliable comparison of theory with observation.
	Important 
	progress in calculations of the relic density 
	of neutralinos in a variety of supersymmetric models 
	has been already made 
\cite{goldberg83,ehnos,krauss83,griest88,gkt,erl90,os91,DreesNojiri,%
	GriestSeckel,GondoloGelmini,an93,lny93,ows,%
	BaerBrhlik,barb,%
	Bottino,leszek,%
	EdsjoGondolo,Darksusy,%
	Ellis-Higgs,EFOS-stau,%
	Belanger:2001fz,%
	Gomez:2000sj,Gomez:2000ck,Lahanas:2000uy,%
	Arnowitt:2001yh,Corsetti:2001yq,%
	Boehm:2000bj,%
	Ellis:2001nx,%
	Belanger:2001am,%
	Nihei:2001qs,%
	Baer:2002fv,%
	Nihei:2002ij,%
	Mizuta:1993qp,%
	Nihei:2002sc,%
	Santoso:2002xu,%
	Bednyakov:2002js}.

	In the early universe neutralinos existed in 
	thermal equilibrium with the cosmic thermal plasma.
	As the universe expanded and cooled, the thermal energy is 
	no longer sufficient to produce neutralinos at an appreciable rate, 
	they decouple and their number density scales with co-moving volume. 
        The sparticles significantly heavier than the LSP decouple at 
	an earlier time and decay into LSPs before the LSPs 
	themselves decouple. 
	Nevertheless there may exist some other next-to-lightest sparticles 
	(NLSPs) which are not much heavier than the stable LSP. 
	The number densities of
	the NLSPs have only slight Boltzmann suppressions with respect to the
	LSP number density when the LSP freezes out of chemical equilibrium
	with the thermal bath.
	Therefore they may still be present in the thermal plasma, 
	and NLSP-LSP and NLSP-NLSP interactions keep 
	the LSP in thermal
	equilibrium resulting in significant reduction of the LSP
	number density and leading to acceptable values even with a rather 
	heavy sparticle spectrum
\cite{EFOS-stau}.
	These {\em coannihilation}\ processes can be particularly 
	important when the LSP-LSP annihilation rate itself is suppressed
\cite{GriestSeckel,Mizuta:1993qp,EdsjoGondolo}.
	Any SUSY particle can be involved in the coannihilation process 
	provided its mass is almost degenerate with the mass of the LSP
\cite{GriestSeckel,Belanger:2001fz}.
	In the low-energy effective MSSM (effMSSM), 
	where one ignores restriction from unification assumptions and 
	investigates the MSSM parameter space at the weak scale
\cite{EdsjoGondolo,Bottino,BKKmodel,Kim:2002cy}
	there is, in principle, no preference for the 
	next-to-lightest SUSY particle.

	The relativistic thermal averaging formalism 
\cite{GondoloGelmini} was extended to include coannihilation processes in 
\cite{EdsjoGondolo}, and was implemented in the DarkSusy code
\cite{Darksusy} 
	for coannihilation of charginos and heavier neutralinos.
	In was found 
\cite{EdsjoGondolo} that for higgsino-like LSP 
	such a coannihilation significantly decreases the relic density 
	and rules out these LSPs from the region of cosmological interest.

  	The importance of the neutralino coannihilation 
	with sfermions was emphasized and investigated for sleptons
\cite{EFOS-stau,Gomez:2000sj}, stops  
\cite{Boehm:2000bj,Ellis:2001nx} and sbottoms
\cite{Arnowitt:2001yh} in the so-called constrained MSSM (cMSSM) 
\cite{leszek,an93,Ellis:2002rp} or in supergravity (mSUGRA) models 
\cite{sugra}.
	The most popular mSUGRA model 
\cite{sugra} 
	has a minimal set of parameters:
$m_0,\ m_{1/2},\ A_0,\ \tan\beta\ {\rm and}\ {\rm sign}(\mu ).$
	Here $m_0$ is the universal scalar mass, 
	$m_{1/2}$ is the universal gaugino mass and 
	$A_0$ is the universal trilinear mass, all evaluated at $M_{\rm GUT}$, 
	$\tan\beta$ is the ratio of Higgs field vacuum expectation values
	and $\mu$ is a Higgs parameter of the superpotential.
	There are strong correlations of sfermion, 
	Higgs boson and gaugino masses in mSUGRA originating
	from unification assumptions. 
	In regions of the mSUGRA parameter space where 
	$\chi$ and ${\tilde \tau}_1$ 
	were nearly degenerate (at low $m_0$), coannihilations could give 
	rise to reasonable values of the relic density even at 
	very large values of $m_{1/2}$, at both low and high $\tan\beta$
\cite{EFOS-stau,Arnowitt:2001yh}. 
	In addition, for large values of the parameter 
	$A_0$ or for non-universal scalar masses, 
	top or bottom squark masses could become nearly degenerate 
	with the $\chi$, so that squark coannihilation processes can become
	important as well
\cite{Boehm:2000bj,Ellis:2001nx}. 
	Therefore due to slepton and squark coannihilation effects, 
	the relic density can reach the cosmologically interesting range of 
	$0.1< \Omega h^2  <0.3$. 

	The influence of coannihilation channels 
	on the LSP proton scalar elastic cross sections
	was considered in supergravity and D-brain models in 
\cite{Arnowitt:2001yh} and in a mSUGRA-like SUSY model for large
	$\tan\beta$  (only for stau coannihilations) in 
\cite{Gomez:2000ck}.

	Having in mind investigation of future 
	prospects for direct and indirect detection of relict LSP
	we follow the most phenomenological view, 
	not bounded by theoretical restrictions from 
	sfermion/gaugino/Higgs mass unifications, etc. 
	To this end we need maximally general and accurate calculations 
	of the relic density within 
	the low-energy effective MSSM scheme (effMSSM)
\cite{Bottino,BKKmodel}.
	The high-level tool for these calculations is the DarkSusy code
\cite{Darksusy}. 
	Unfortunately the code calculates only 
	neutralino with next neutralino(s) and chargino 
	coannihilations (NCC), 
	which is not sufficient, when 
	neutralino-slepton coannihilation (SLC) and 
	neutralino-squark (SQC) coannihilation are claimed to be dominant
\cite{EFOS-stau,%
	Gomez:2000sj,%
	Boehm:2000bj,%
	Ellis:2001nx,%
	Arnowitt:2001yh}.

	Contrary to the majority of previous papers (see for example
\cite{EFOS-stau,Ellis:2001nx,Lahanas:2000uy,Gomez:2000sj,Nihei:2002sc}), 
	aimed mostly at discovery and demonstration 
	of the importance or 
	dominance of some specific coannihilation channels,
	our main goal is the investigation of the interplay
	between different coannihilation channels
	as well as their consequences for detection of CDM. 	 
	To this end a comparative study of NCC and SLC channels, 
	exploration of relevant changes in the relic density 
 	and investigation of their consequences for detection of  
	CDM particles were performed in the effMSSM
	in our previous paper 
\cite{Bednyakov:2002js}.
	The present paper extends our investigations 
\cite{Bednyakov:2002js} to the 
	neutralino-stop and neutralino-sbottom coannihilations 
	and completes our consideration of the subject. 
	For this purpose we combined our previuos code
\cite{BKKmodel} with the DarkSusy code 
\cite{Darksusy} and codes of 
\cite{EFOS-stau,Ellis:2001nx}, 
	which allows for the first time 
	comparative study of relevant coannihilation channels 
	(NCC, SLC, SQC) in the low-energy effMSSM.

\section{The \lowercase{eff}MSSM approach}
	As free parameters in the effMSSM, we use 
	the gaugino mass parameters $M_1, M_2$,
	the entries to the squark and slepton mixing matrices 
	$m^2_{\tilde Q}, m^2_{\tilde U}, m^2_{\tilde D}, 
	 m^2_{\tilde R}, m^2_{\tilde L}$ 
	for the 1st and 2nd generations and 
	$m^2_{\tilde Q_3}, m^2_{\tilde T}, m^2_{\tilde B}, 
	 m^2_{\tilde R_3}, m^2_{\tilde L_3}$
	for the 3rd generation, respectively; 
	the 3rd generation trilinear soft couplings $A_t , A_b , A_\tau$; 
	the mass $m_A$ of the pseudoscalar Higgs boson, 
	the Higgs superpotential parameter $\mu$, and $\tan\beta$.
	To reasonably reduce the parameter space we assumed 
$ m^2_{\tilde U} = m^2_{\tilde D} = m^2_{\tilde Q}$;
$ m^2_{\tilde T} = m^2_{\tilde B} = m^2_{\tilde Q_3}$;
$ m^2_{\tilde R} = m^2_{\tilde L}$;
$ m^2_{\tilde R_3} = m^2_{\tilde L_3}$
	and have fixed $A_b = A_{\tilde \tau} = 0$
\cite{BKKmodel}. 
	The third gaugino mass parameter $M_3$ defines the 
	mass of the gluino in the model and is 
	determined by means of the GUT assumption $M_2 = 0.3\, M_3$.
	The remaining parameters defined our effMSSM 
	parameter space and were scanned randomly within the
	following intervals: 
\begin{eqnarray*} 
- 1 \tev < M_1 < 1\tev, \quad 
-2\tev < M_2 , \mu , A_t < 2\tev, \quad 
1.5<\tan\beta < 50, \\  
50\gev < M_A < 1000\gev, \quad
10 \gev^2 < m^2_{\tilde Q}, m^2_{\tilde L}, 
m^2_{\tilde Q_3}, m^2_{\tilde L_3} < 10^6\gev^2.
\end{eqnarray*}
	We have included the current experimental 
	upper limits on sparticle masses
	as given by the Particle Data Group 
\cite{pdg}. 
	The limits on the rare $b\rightarrow s \gamma$ decay 
\cite{flimits} following 
\cite{BerBorMasRi} have also been imposed. 
	The calculations of the neutralino-nucleon cross sections and direct 
	and indirect detection rates follow the description given in 
\cite{jkg,BKKmodel}.

	The number density is governed by the Boltzmann equation
\cite{GondoloGelmini,EdsjoGondolo} 
\begin{equation} \label{boltzmann} 
{d n \over dt}+ 3 H n = - {\langle\sigma v\rangle} (n^2 - n_{\rm eq}^2 ) 
\end{equation} 
      	with $n$ either being the LSP number density if there are no other 
      	coannihilating sparticles, or 
	the sum over the number densities of all coannihilation partners.
	The index ``eq'' denotes the corresponding equilibrium value.
	To solve the Boltzmann equation 
(\ref{boltzmann}) one needs to evaluate the thermally averaged neutralino 
	annihilation cross section ${\langle \sigma v \rangle}$. 
	Without coannihilation processes ${\langle \sigma v \rangle}$ 
	is given as the thermal average of the LSP 
	annihilation cross section $\sigma_{\chi\chi}$
	multiplied by relative velocity $v$ of the annihilating LSPs
\begin{equation} \label{sigmano} 
{\langle \sigma v \rangle} = {\langle \sigma_{\chi\chi} v \rangle} , 
\end{equation}
	otherwise it is determined as
	${\langle \sigma v \rangle} = 
	{\langle \sigma_{\rm eff} v \rangle}$, where  
	the effective thermally averaged cross section
	is obtained by summation over coannihilating particles
\cite{GondoloGelmini,EdsjoGondolo}
\begin{equation} 
{\langle \sigma_{\rm eff} v \rangle} = 
\sum_{ij} {\langle \sigma_{ij}v_{ij} \rangle}
	 {n^{\rm eq}_i \over n^{\rm eq} }{n^{\rm eq}_j 
\over n^{\rm eq}}.
\label{sigmaeff} 
\end{equation}
      	If $n_0$ denotes the present-day 
	number density of the relics, the relic density  is given by
\begin{equation} \label{finres} 
\Omega = {m_\chi n_0 \over \rho_{c}}.
\end{equation}

	For each point in the MSSM parameter space (MSSM model) 
	we have evaluated the relic density of the LSP 
	ignoring any possibility of coannihilation (IGC), taking into 
	account only neutralino-chargino (NCC), slepton (SLC), 
	or squark (SQC) coannihilations separately, 
	and including all of the coannihilation channels (ACC).
	To this end DarkSusy procedures of 
	${\langle \sigma_{\rm eff} v \rangle}$
	evaluation and solution of Boltzmann equation were implemented
	in our former code 
\cite{BKKmodel}.
	Coannihilations with two-body 
	final states that can occur between neutralinos, charginos,
	sleptons, stops and sbottoms,
	as long as their masses are $m_i<2\mchi$, were included.
	The Feynman amplitudes for NCC, SLC and stop coannihilations
	were taken from DarkSusy 
\cite{Darksusy}, 
\cite{progfalk,EFOS-stau}, and  
\cite{progsantoso,Ellis:2001nx}, respectively. 
	The amplitudes for the sbottom coannihilation  	
	were obtained on the basis of the stop amplitudes from  
\cite{progsantoso,Ellis:2001nx}.
	As in 
\cite{Bednyakov:2002js},
	the ${\langle \sigma_{\rm eff} v \rangle}$ 
	and ${\Omega h^2 }$ were calculated following 
	the relevant DarkSusy routines 
\cite{Darksusy} to which the codes  
\cite{progfalk,EFOS-stau}, and  
\cite{progsantoso,Ellis:2001nx} were added
	in a way that guarantees the correct inclusion of 
	SLC and SQC.

	In the case where all squarks, sleptons, neutralinos and 
	charginos are substantially heavier than the LSP 
	($m_i>2\mchi$) and there is no possible coannihilations, 
	the relic density
	$\Omega h^2 = \ACC = \NCC = \SLC = \SQC$ is equal to 
	$\Omega_\chi h^2$ obtained without any coannihilations. 
	When, for example, at least one of the 
	coannihilation channels (NCC, SQC or SLC) is indeed relevant, 
	the $\IGC$ (ignoring coannihilation) is calculated with 
\begin{equation} 
{\langle \sigma_{\rm eff} v \rangle}_{\rm IGC} = 
{\langle \sigma_{\chi\chi} v \rangle} 
 \left( n_\chi^{\rm eq} \over n^{\rm eq} \right)^2,
\end{equation}
	where $n^{\rm eq}$ includes {\em all}\ 
	open coannihilation channels. 
	This formula (generalized also for NCC, SLC and SQC) 
	allows a comparative study of 
	all coannihilation channels, always 
	leading to a smaller value for the relic density, $\COA / \IGC < 1$. 
	Here $\COA$ is 
	a common notation for $\ACC$,  $\NCC$, $\SQC$ or $\SLC$. 
	We assume $0.1< \Omega h^2  < 0.3$ 
	for the cosmologically interesting region
\cite{acc}.

\section{Results and Discussions}
\subsection{Coannihilation effects in the relic density}
\begin{figure}[h] 
\begin{picture}(100,110)
\put(-10,-58){\includegraphics{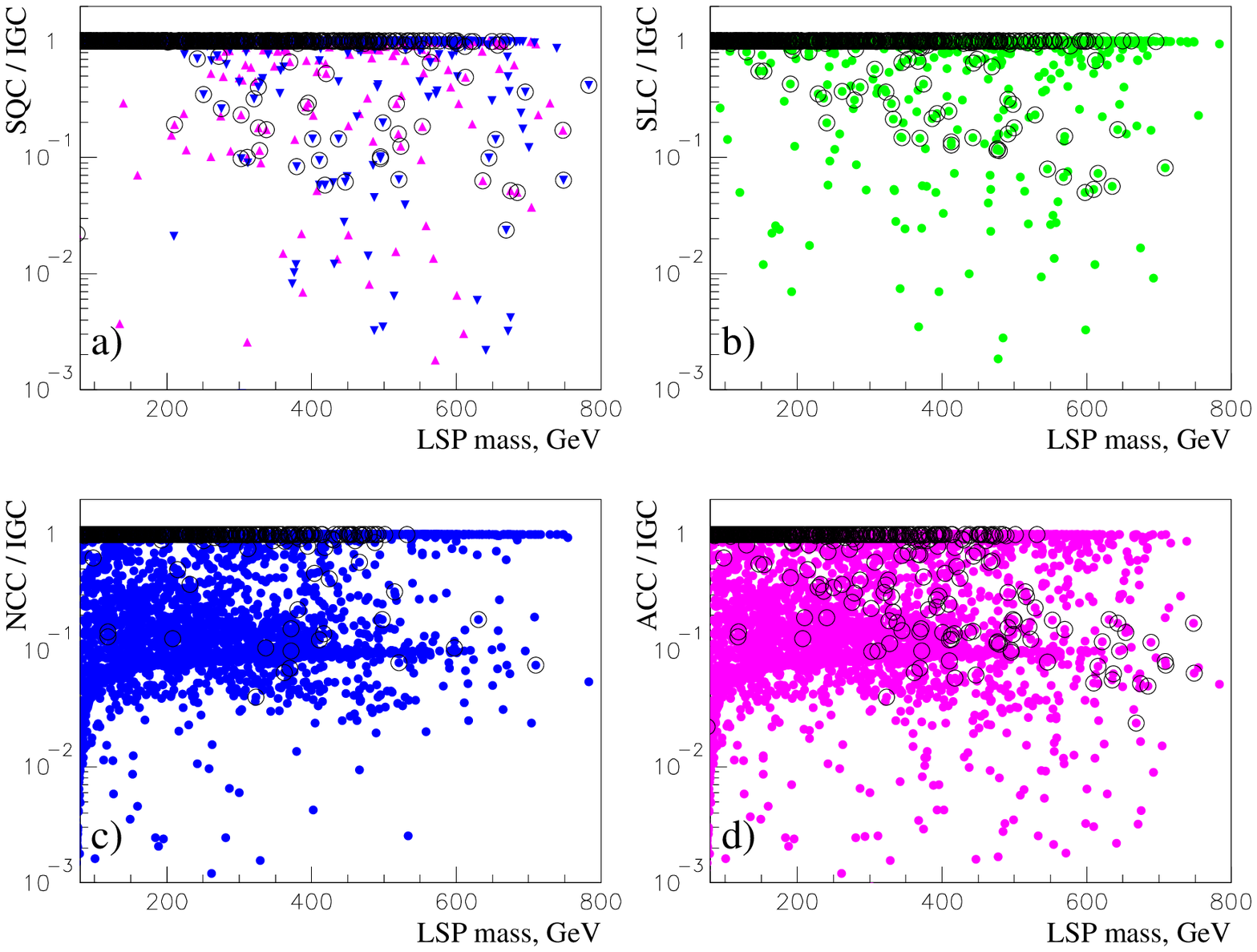}}
\end{picture}
\caption{Effects of  squark-neutralino (SQC), 
	            slepton-neutralino (SLC), and 
	neutralino-chargino\-(neutralino) (NCC) coannihilations in effMSSM. 
	Panels a)~--~d) display ratios 
	$\SQC / \IGC$, 
	$\SLC / \IGC$, 
	$\NCC / \IGC$, and 
	$\ACC / \SLC$ for the case when all coannihilations are included.
	The maximal reduction factors for all channels (NCC, SQC, and SLC) 
	are of the order of $10^{-3}$.
	Encircled points 
	mark the cosmologically interesting relic density $0.1< \COA < 0.3$. 
	In panel a) up-going triangles correspond to stop coannihilations
	and down-going triangles correspond to sbottom coannihilations. 
	The stop and sbottom contributions are seen to be equal. 
\label{A2I-f01}}
\end{figure}

\begin{figure}[ht] 
\begin{picture}(100,100)
\put(-10,-65){\includegraphics{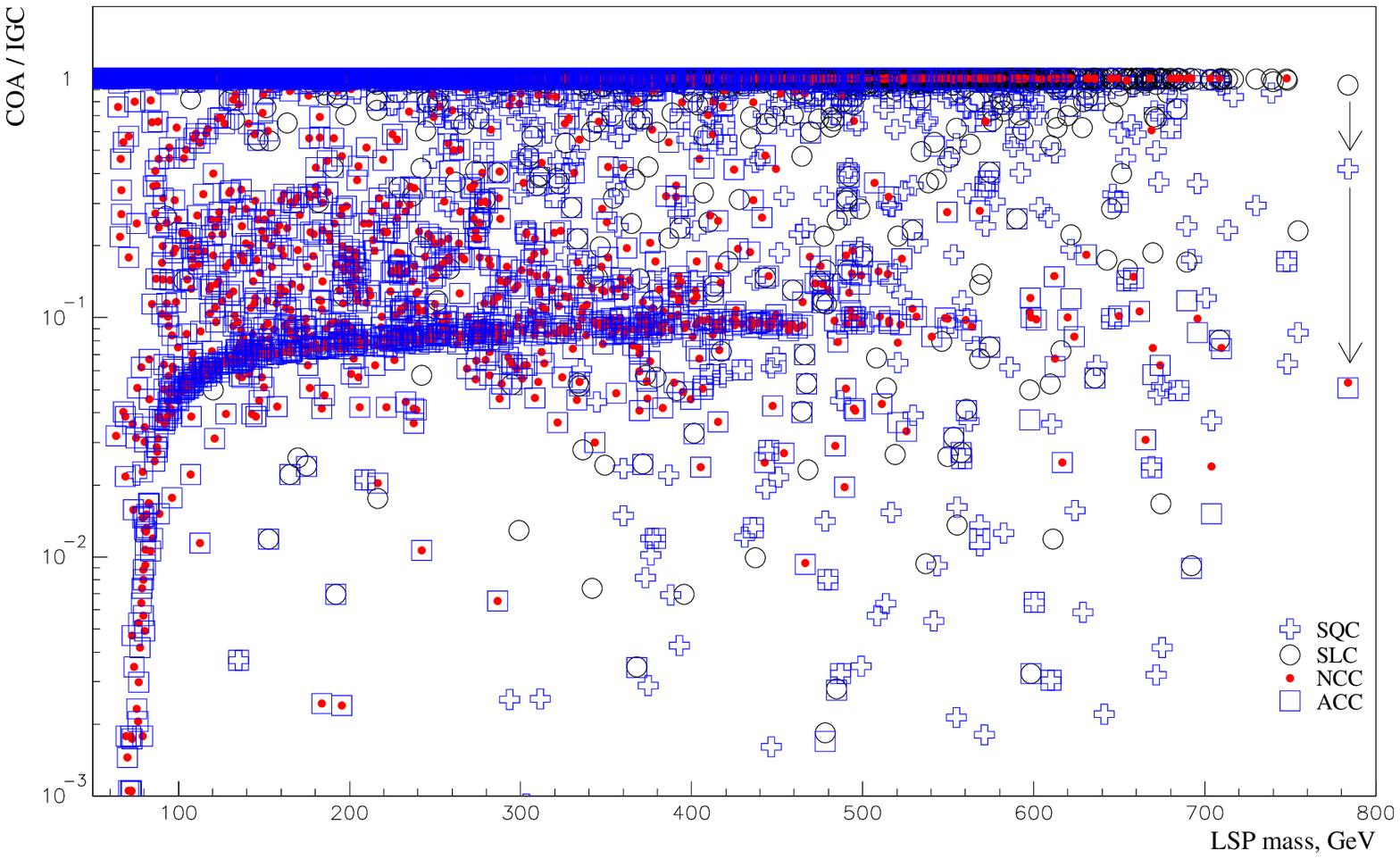}}
\end{picture}
\caption{The same as in 
Fig.~\ref{A2I-f01}, but plotted together.
 	Here 
	$\SQC / \IGC$, 
	$\SLC / \IGC$, 
	$\NCC / \IGC$, and 
	$\ACC / \SLC$  are marked 
	with crosses, circles, dots and squares, respectively. 
	Therefore, a square filled with a cross, circle, or dot 
	depicts a model that is affected only by SQC, SLC, or NCC, 
	respectively, while 
	any other coannihilation channel gives a negligible contribution.
	Such a situation takes place for the majority of models, 
	but there are some (quite few) models, given by empty squares, 
	for which at least two coannihilation channels are relevant. 
	For example, arrows on the right side of the figure 
	demonstrate how reduction of RD proceeds:
	SLC gives no effect ($\SLC / \IGC=1$),
	SQC reduces RD with a factor $\SQC / \IGC\approx 0.4$,
	and finally NCC gives the main contribution to RD suppression,
	$\ACC / \SLC \approx\NCC / \IGC \approx 0.04$ 
	(the square nearly coincides with the dot).   
\label{A2I-f02}}
\end{figure}

\begin{figure}[ht] 
\begin{picture}(100,177)
\put(-8,-28){\includegraphics{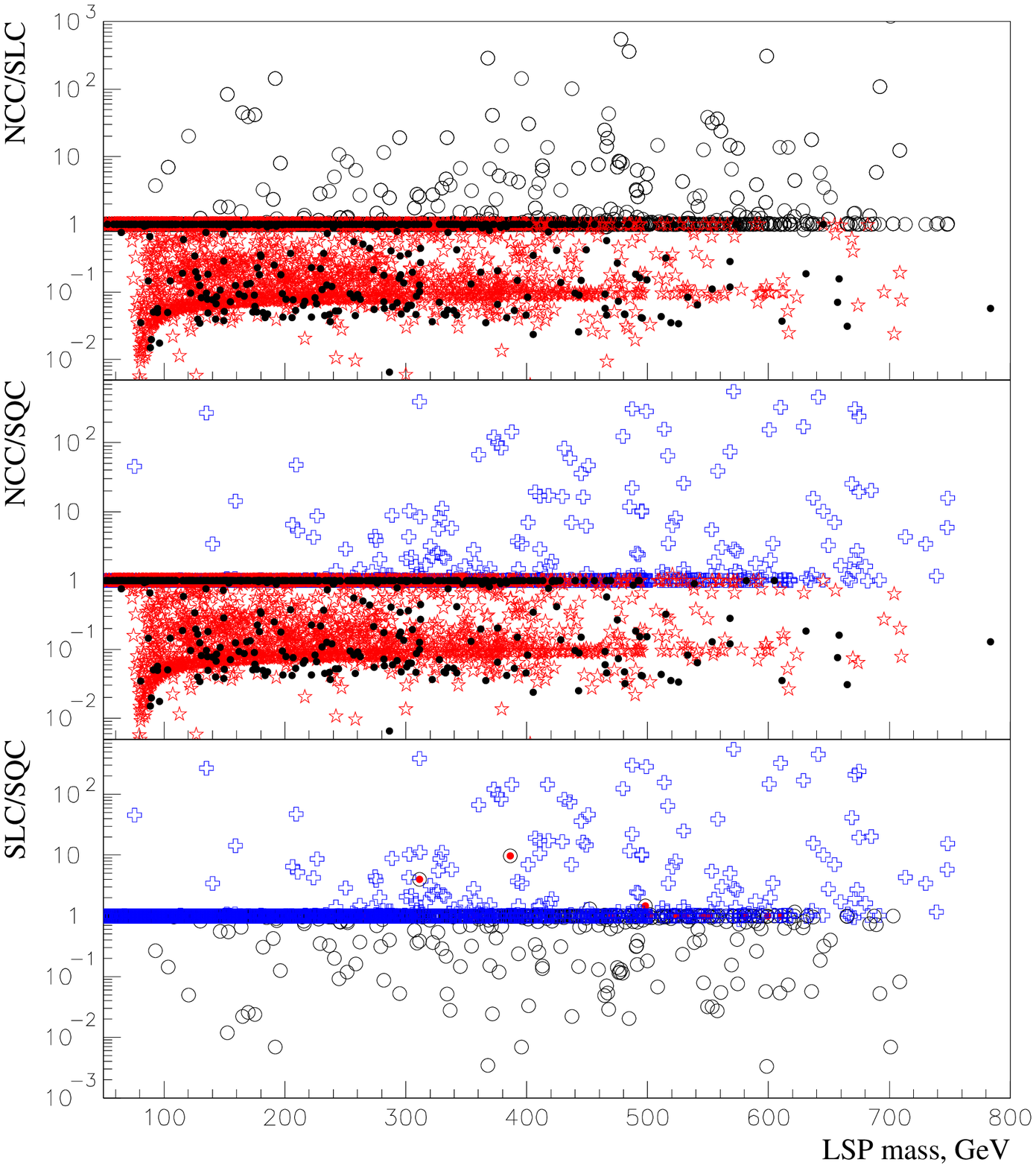}}
\end{picture}
\caption{Ratios $\NCC / \SLC$, $\NCC / \SQC$ and $\SLC / \SQC$ versus $\mchi$.
	An open circle indicates that the $\tilde\tau$ is the NLSP, 
	the star means that the light chargino $\tilde\chi^\pm$  
	is the NLSP, small filled circle marks the model where the 
	second-lightest neutralino $\tilde\chi_2$ is the NLSP. 
	An open cross indicates that the stop $\tilde t$ or the 
	sbottom $\tilde b$ is the NLSP.
	One sees, for example, that if 
	$\tilde\chi_2$ or $\tilde\chi^\pm$ is the NLSP,  
	the NCC necessarily dominates ($\NCC / \SLC<1$, or $\NCC / \SQC< 1$),
	while $\tilde t$ or $\tilde b$ being the NLSPs always
	leads to dominant SQC 
	($\NCC / \SQC > 1$, or $\SLC / \SQC> 1$).
	The same is in general true for SLC.
	Circles with dots inside depict models where the NLSP is the stau, 
	but SLC does not dominate. 
\label{NC2SF-f03}}
\end{figure}

\begin{figure}[h] 
\begin{picture}(100,130)
\put(-5,-45){\includegraphics{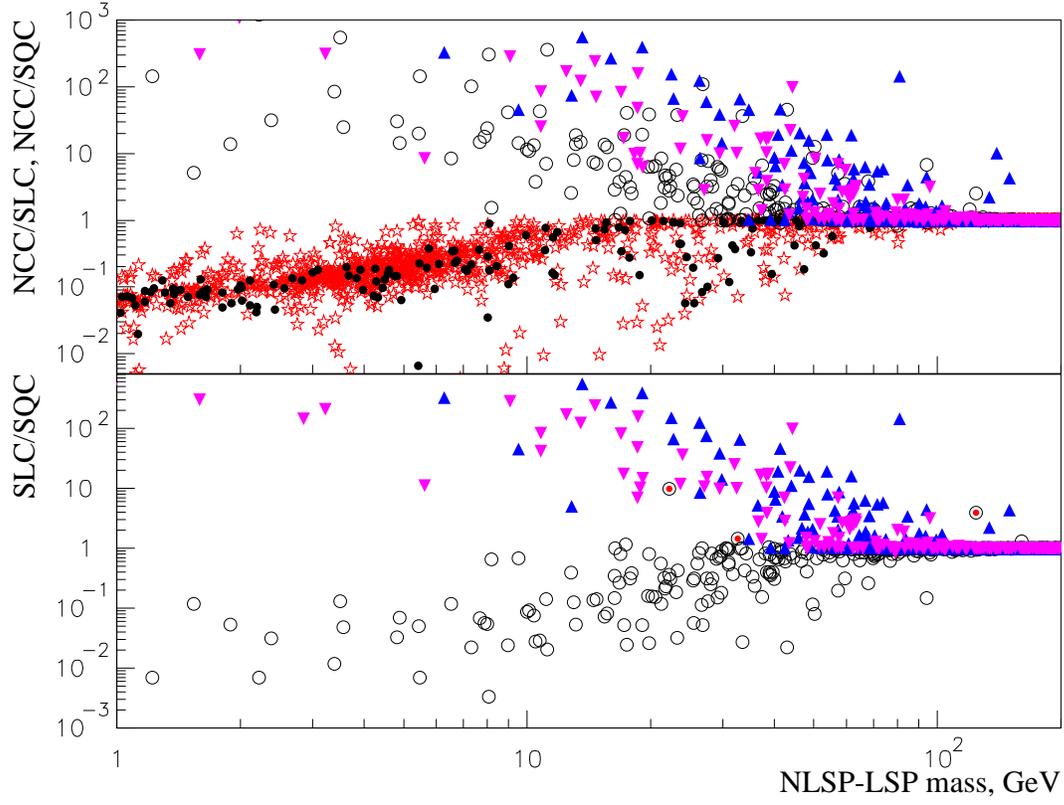}}
\end{picture}
\caption{The same as in 
Fig.~\ref{NC2SF-f03}, but versus $m^{}_{\rm NLSP} - \mchi$. 
	Up-going triangles correspond to stops
	and down-going triangles correspond to sbottoms. 
\label{MD-f04}}
\end{figure}

\begin{figure}[ht] 
\begin{picture}(100,140)
\put(-8,-55){\includegraphics{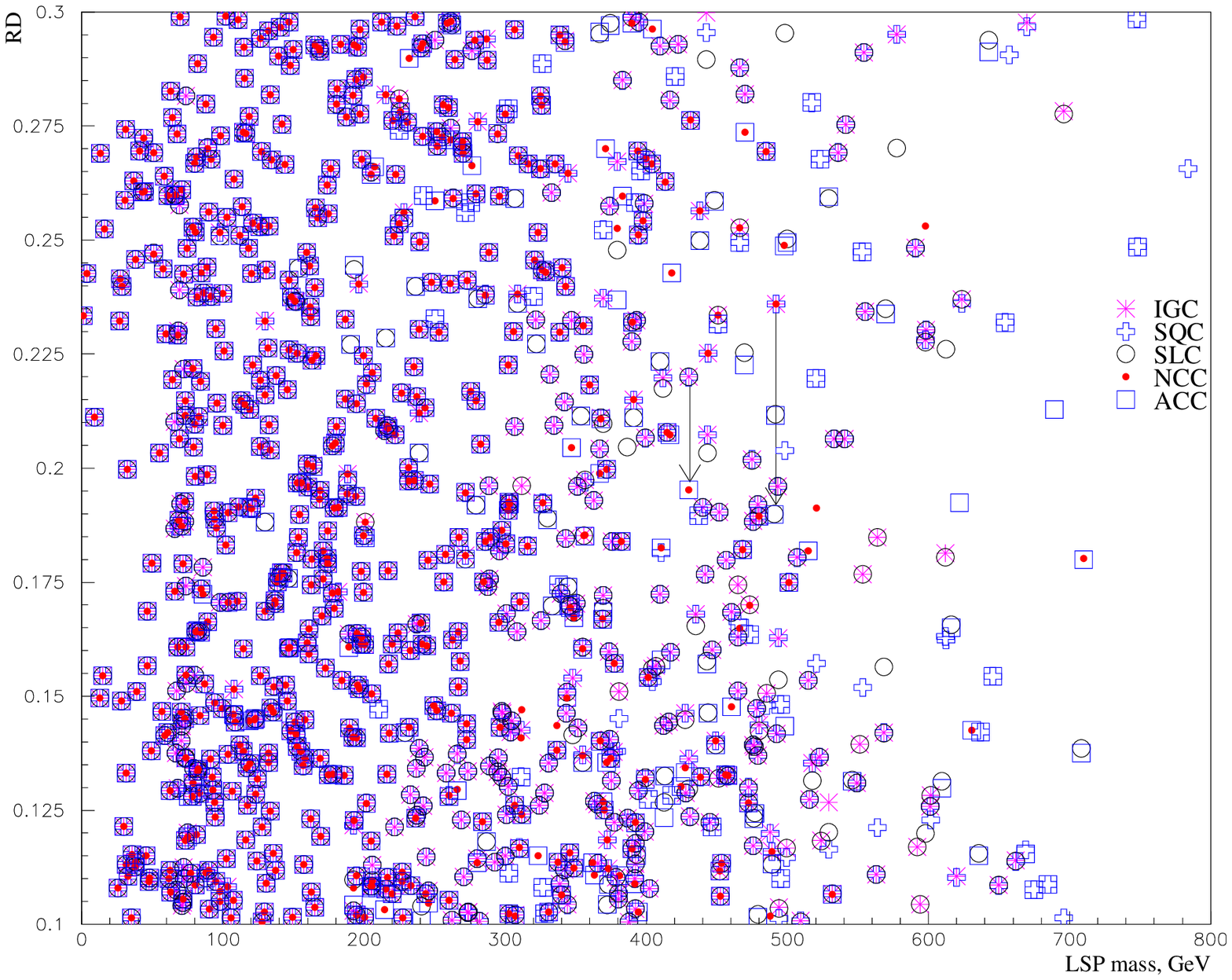}}
\end{picture}
\caption{Illustration of the shifting of effMSSM models into 
	and out of 
	the cosmologically interesting range 
	$0.1< \COA < 0.3$ due to NCC, SQC and SLC. 
	Relic densities $\IGC$, $\SQC$, $\SLC$, $\NCC$ and $\ACC$ 
	are marked with stars, crosses, circles, small dots, and squares,
	respectively. 
	Therefore, a superposition of all those symbols corresponds 
	to a model which is totally untouched by coannihilation. 
	A star-crossed circle marks a model which is unaffected by SLC and SQC 
	($\SLC=\SQC=\IGC$), but is shifted down due to NCC. 
	If the corresponding $\ACC$ (which is equal to $\NCC$) 
	remains within this range, it is still present in the figure 
	below this star-crossed circle
	as an square with a black dot inside (see short arrow).
	By analogy, a square with a circle inside gives 
	a model which is shifted into the region due to SLC only
	($\ACC=\SLC$), and if the corresponding $\IGC=\NCC=\SQC$
	is also in the cosmologically viable range, 
	it is located above the symbol 
	as a crossed star with a dot inside (see long arrow).
	Quite a large number of models are shifted out of the range  
	$0.1< \Omega h^2<0.3$ due to NCC (star-crossed circles).
\label{ACC-f05}}
\end{figure}

\begin{figure}[h] 
\begin{picture}(100,75)
\put(-4,-70){\includegraphics{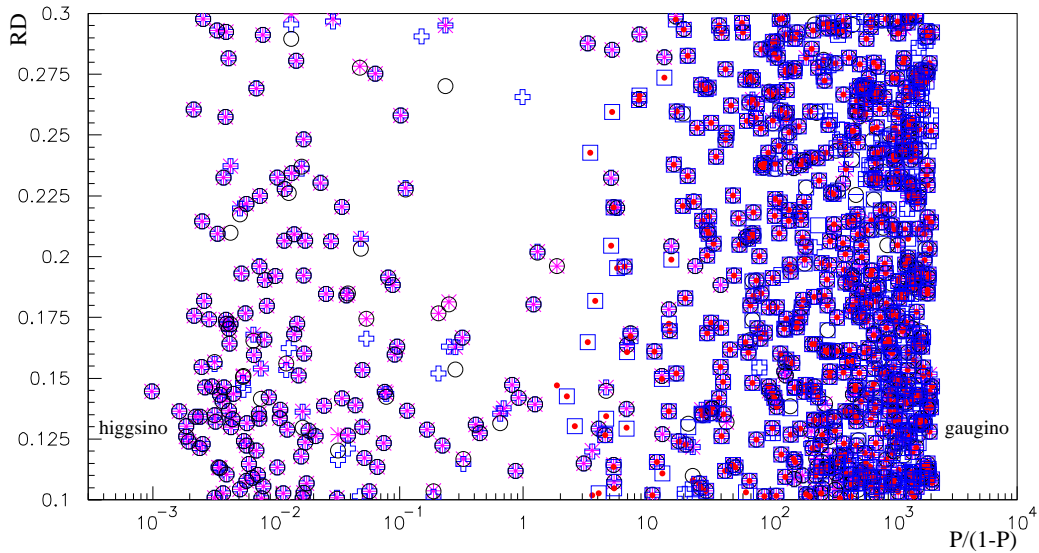}}
\end{picture}
\caption{Variation of relic density against gaugino fraction.
	As in 
Fig.~\ref{ACC-f05}, RD $\IGC$, $\SQC$, $\SLC$, $\NCC$ and $\ACC$ 
	are marked with stars, crosses, circles, small dots, and squares,
	respectively. 
	The NCC reduces the RD especially for 
	models with higgsino-like LSPs and shifts
  	these models out of cosmological interest. 
	The joint effect of NCC, SQC and SLC leaves 
	only LSPs with  ${\cal P} > 0.67$
	in the cosmologically interesting region.
\label{GF-f06}} 
\end{figure}

	We performed our calculations in the effMSSM approach given above 
	and the results 
	for the neutralino relic density (scatter plots) are presented in  
	Figs.
\ref{A2I-f01}--\ref{GF-f06}. 
	The reduction effect on the relic density (RD) produced by 
	SQC, SLC, NCC and ACC is shown in 
Fig.~\ref{A2I-f01} as a ratio $\COA / \IGC$.
	On the basis of our sampling (50000 models tested)
	the maximum RD suppression factor for the NCC and SLC channels 
	is of the order of $10^{-3}$.
	Almost the same maximal suppression is found 
	for the squark coannihilation channels.
	These results depend on the 
	experimental limits imposed on the second-lightest neutralino, 
	chargino and slepton stop and sbottom masses.
	If there were no limits 
	on their masses, the factor of relative RD reduction due to NCC  
	could reach a maximum value of $10^{-5}$ for 
	models with $\mchi \approx 40\gev$ 
\cite{Bednyakov:2002js}.
	The current experimental limits for 
	$m_{\tilde \tau}$, 
	$m_{\tilde\chi^\pm}$, $m_{\tilde t}$, and
	$m_{\tilde b}$ are 80--90$\gev$ 
\cite{pdg},
	and therefore the critical LSP mass that enables 
	non-negligible NCC, SLC, and SQC contributions 
	is also of the same order ($\mchi \ge 80\gev$).

	From panel a) of the figure 
	one can conclude that stop (up-going triangles) 
	and sbottom (down-going triangles) equally contribute
	to reduction of the RD due to coannihilations. 	

	The encircled symbols 
	depict some kind of ``constructive'' 
	reduction, when due to the coannihilations the relic density falls 
	into the cosmologically interesting region $0.1< \COA < 0.3$. 
  	Other points present the cases when coannihilations 
	too strongly reduce the relic density. 
	One can see that NCC plays the main role in
	``destructive'' reduction of RD,  
	these channels reduce the maximal number of models 
	from the cosmologically interesting region
\cite{EdsjoGondolo,Bednyakov:2002js}.
	Despite this, 
	all coannihilation channels contribute equally much 
	in the ``constructive'' reduction of RD 
	(there are almost the same 
	number of the circled points in  panels a)--c)).
	Therefore for our random sampling in the effMSSM 
	the NCC, SLC and SQC are indeed relevant 
	($\COA/\IGC< 0.95$) for 2--4\% of models if one 
	assumes $0.1 < \ACC < 0.3$, and 
	the NCC is indeed relevant for about 30\% of models
	if this constraint is relaxed 
(Fig.~\ref{A2I-f01}). 

	From 
Fig.~\ref{A2I-f02} one can see that 
	mainly only one of the coannihilation channels (NCC, SQC, or SLC)
	dominates in the reduction of RD.
	The other channels of coannihilation in general play no role or 
	lead only to a much smaller further reduction
\cite{Bednyakov:2002js}.

  	Although other coannihilation processes besides NLSP-LSP can 
	in principal be also open 
	(including LSP coannihilation with the next-to-NLSP (NNLSP) 
	and next-to-NNLSP, as well as NLSP-NLSP coannihilations, etc), 
Fig.~\ref{NC2SF-f03} allows the
	conclusion that	the dominant coannihilation channel is 
	defined by the type of the NLSP.	
	If the next neutralino $\tilde\chi_2$ or 
	chargino $\tilde\chi^\pm$ is the NLSP, then  
	NCC indeed dominates. 
	The SQC dominates when NLSP is the stop or the sbottom.
	The stau $\tilde\tau$ (or another slepton) being the 
	NLSP entails a dominant SLC effect.

	Nevertheless, contrary to the NCC$+$SLC case  
\cite{Bednyakov:2002js},
	there are (very few) models where the stau is the NLSP, but  
	masses of the stau, the stop and the sbottom appear 
	by accident almost the same and all of these sparticles
	participate in the coannihilation with the LSP and each other.
	As a result, strongly interacting coannihilation
	channels with squarks produce a larger reduction as compared to the
	SLC despite the fact that the stau is indeed the NLSP.
	The bottom panels of 
Figs.~\ref{NC2SF-f03} and~\ref{MD-f04} display
	an example of such models (circle with a dot inside),
	in which the stau is the NLSP but the SLC contribution is smaller
	than the contribution of SQC ($\SLC / \SQC > 1$). 

	Figure
\ref{MD-f04} shows that for all coannihilation channels
	maximal RD reduction factors (less than 0.01)
	occur for mass differences 
	$m^{}_{\rm NLSP} - m^{}_{\rm LSP} \le 20\gev$.
	Mass difference $m^{}_{\rm NLSP} - m^{}_{\rm LSP} \le 5\gev$
	plays a significant role in RD reduction 
	for SLC and mostly for NCC.
	In contrast with NCC and SLC, 
	SQC can produce the same RD reduction effect with
	larger mass difference between squarks and the LSP 
	($m^{}_{\tilde q} - m^{}_{\rm LSP}\approx 150\gev$)
	due to the possibility of coannihilation via 
	the strong interaction.

	In the case of SQC the small 
	mass difference between the coannihilating stop and sbottom quarks 
	($m^{}_{\tilde t} - m^{}_{\tilde b}$ 
	instead of $m^{}_{\tilde q} - m^{}_{\rm LSP}$)
	can produce a dominant effect in RD reduction
	(triangles at $m^{}_{\tilde q} - m^{}_{\rm LSP}\ge 100\gev$ in 
Fig.~\ref{MD-f04} illustrate this possibility).

	Although we have set
	the coannihilation opening threshold of $m_i = 2\mchi$, 	
	for NCC and SLC channels, 
	relevant effects occur if the mass difference between 
	the coannihilation partner and the LSP is within 15\%. 
	This is in agreement with previous considerations 
\cite{GriestSeckel,%
	EFOS-stau,%
	Lahanas:2000uy,%
	EdsjoGondolo,%
	Corsetti:2001yq,%
	Gomez:2000sj,%
	Ellis:2001nx,%
	Arnowitt:2001yh}.
	It was found that for SQC 
	the relevant effects occur if the mass difference between 
	the coannihilating squark and the LSP is within 50\%
	(in general agreement with 
\cite{Ellis:2001nx,Santoso:2002xu}). 

	In 
Fig.~\ref{ACC-f05} all calculated 
	relic densities ($\IGC$, $\SQC$, $\SLC$, $\NCC$ and $\ACC$)
	are depicted in the cosmologically interesting region  
	$0.1< \COA < 0.3$.
	A large number of models (mostly with $\mchi \le 250\gev$) 
	are completely unaffected by any kind of coannihilation.
	When at least one of the coannihilation channels is relevant, 
	the RD decreases and some cosmologically unviable models with  
	$\IGC > 0.3$ enter the cosmologically interesting range
	$0.1< \COA < 0.3$,
	due to NCC (squares with a dot inside), 
	SLC (squares with circles inside),
	SQC (squares with crosses inside),
	or due to joint contribution of NCC, SQC, or/and SLC
	(empty squares).

	There are also models which fall within 
	the less interesting region ($\COA < 0.1$).	
	The largest amount of models are shifted out 
	due to NCC (star-crossed circles),
	and a relatively small amount of models is shifted out
	due to SLC (crossed stars with a dot inside), 
	SQC (circles with a star and a dot inside), 
	both NCC and SLC (crossed stars).
	There are cosmologically interesting LSPs 
	within the full mass range $20\gev < \mchi < 720\gev$
(Fig.~\ref{ACC-f05}) accessible in our scan  whether or not
 	coannihilation channels are included.

	Cosmologically interesting LSPs occur with arbitrary compositions
	when coannihilations are ignored 
(Fig.~\ref{GF-f06}), the inclusion of NCC rules out
	all the models with higgsino-like LSPs
	(star-crossed circles), SLC and SQC further tends to rule
	out LSPs with mixed composition, so that only LSPs with 
	${{\cal P}} > 0.67$ 
	(there are no squares for $\frac{{\cal P}}{1-{\cal P}}<2$)
	remain as dominant CDM candidates.
	In general our estimations 
(Fig.~\ref{GF-f06}) are in accordance with previous considerations
\cite{EFOS-stau,EdsjoGondolo},
	as far as they can be compared with these less general treatments.

\subsection{Coannihilation effects in the detection rates}
\begin{figure}[h] 
\begin{picture}(100,92)
\put(-10,-68){\includegraphics{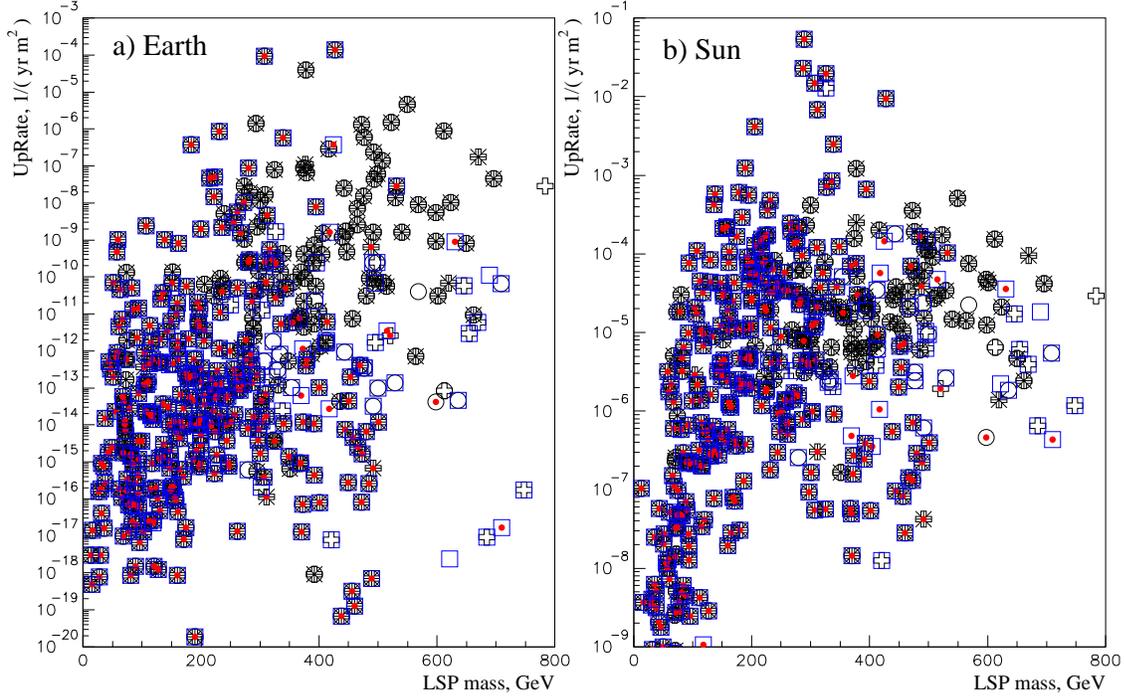}}
\end{picture}
\caption{Indirect detection rate for upgoing muons 
	from $\chi\chi$ annihilation in the Earth (a) and the Sun (b). 
	As in 
Fig.~\ref{ACC-f05}, 
	stars, crosses, circles, small dots, and squares  correspond to
	$0.1<\IGC, \SQC, \SLC, \NCC, \ACC<0.3$, respectively. 
	NCC decreases the detection rates
	for models with $\mchi \ge 400\gev$.
\label{IND-f07}}
\end{figure}

\begin{figure}[h] 
\begin{picture}(100,94)
\put(-10,-70){\includegraphics{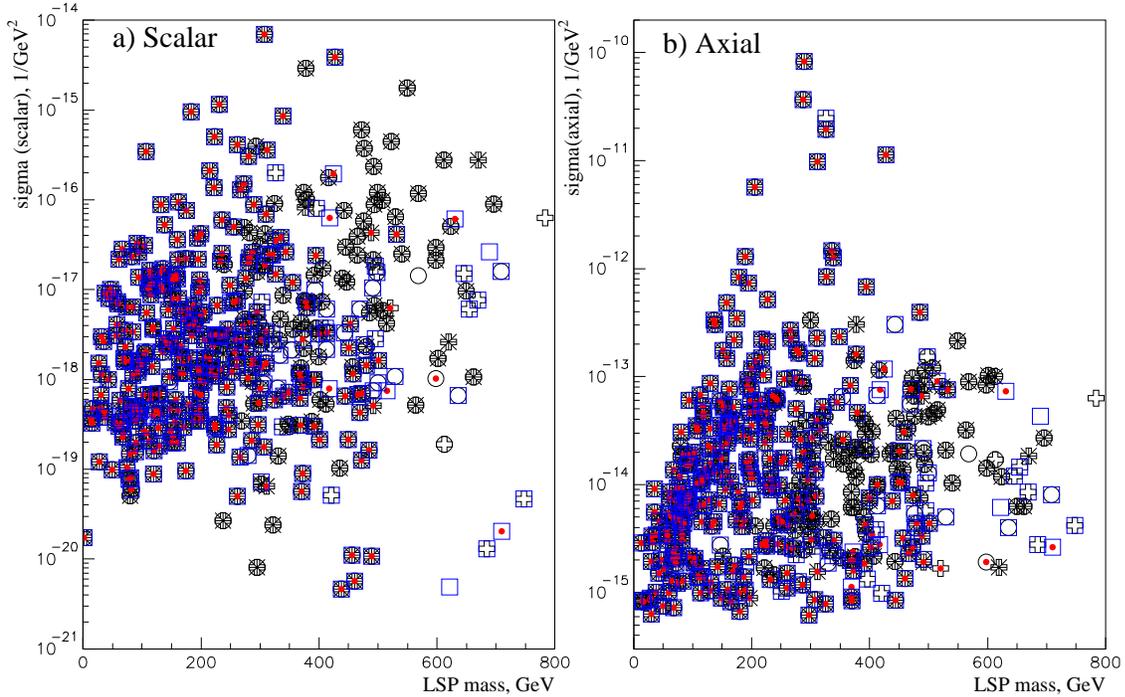}}
\end{picture}
\caption{Neutralino-proton scattering cross sections for 
	scalar (spin-independent) interaction
	(a) and axial (spin-dependent) interaction (b). 
	The notations are as in 
Fig.~\ref{IND-f07}.	
\label{CS-f08}}
\end{figure}

\begin{figure}[h] 
\begin{picture}(100,80)
\put(-10,-75){\includegraphics{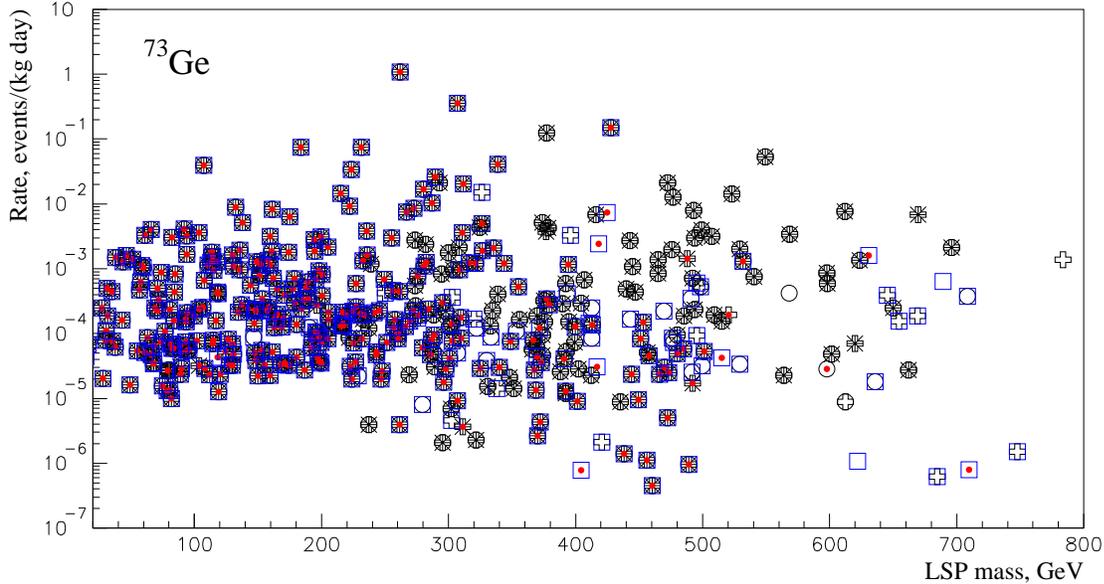}}
\end{picture}
\caption{Event rate for direct neutralino detection in a 
	$^{73}$Ge detector. 
	As in 
Fig.~\ref{ACC-f05}, 
	stars, crosses, circles, small dots, and squares  correspond to
	$0.1<\IGC, \SQC, \SLC, \NCC, \ACC<0.3$, respectively. 
	NCC decreases the maximal event 
	rates for models with $\mchi \ge 500\gev$, 
	but the models with smaller LSP mass 
	are unaffected by the coannihilations.
\label{RGe-f09}}
\end{figure}

	Now we consider the influence of all possible coannihilation channels 
	(NCC, SQC and SLC) on prospects for  
	indirect and direct detection of CDM neutralinos. 
	The results of our calculations (scatter plots) 
	for cold dark matter observables are presented in Figs.
\ref{IND-f07}--\ref{RGe-f09}.
	We compare the rate predictions for cosmologically interesting 
	LSPs when the RD is evaluated with or without any of coannihilation
	channels taken into account. 
	We see 
(Fig.~\ref{ACC-f05}) that in most models with $\mchi \le 250\gev$ 
	the RD is unaffected  by SQC, SLC and NCC, 
	mostly because the difference  $m_{\rm NLSP} - \mchi$ 
	is too large to yield significant effects, 
	therefore the corresponding detection rates are not influenced
	(depicted in the figures as a square 
	filled with a star, a cross and a dot simultaneously). 

Figure~\ref{IND-f07} displays the expected indirect detection rates 
	for upgoing muons produced in the Earth by neutrinos 
	from decay products of $\chi\chi$ annihilation 
	which takes place in the core of the Earth or of the Sun.

	For $\chi\chi$ annihilation in the Earth 
	upgoing muon detection rates merely lie within the range 
	$ 10^{-19} \hgev < \Gamma^\mu <  5 \cdot 10^{-5}\hgev $ 
	as long as $\mchi \le 250\gev$. 
	When $\mchi \ge 250\gev$, some of the models with 
	$0.1<\IGC < 0.3$ are removed 
	from the cosmological interesting range ($\COA < 0.1$;
 Fig.~\ref{ACC-f05}) mainly due to NCC
(Fig.~\ref{IND-f07}).  
	Other models with $\IGC > 0.3$ are shifted into this region
(Fig.~\ref{ACC-f05} and 
 Fig.~\ref{IND-f07}) mainly due to SLC and SQC. 
	For $\mchi \ge 250\gev$ one finds 
	$ 10^{-19}\hgev < {\Gamma^\mu}_{\rm ACC} < 5\cdot 10^{-7}\hgev$
	when the RD is evaluated with 
	coannihilations taken into account, and 
	$10^{-19}\hgev < {\Gamma^\mu}_{\rm IGC} <4\cdot 10^{-6}\hgev$ 
	when coannihilations are neglected.
	The large values of the detection rates of
	$\chi\chi$ annihilation in the Earth are decreased 
	(from $10^{-5}\hgev$ to $10^{-8}\hgev$) only for heavy LSPs, 
	$\mchi > 450\gev$, in accordance with the fact that  
	the corresponding models are removed from the cosmologically 
	interesting range.
	The SQC does not significantly change this conclusion obtained in 
\cite{Bednyakov:2002js} for SLC and NCC.

	In the case of indirect detection of upgoing muons from $\chi\chi$ 
	annihilation in the Sun one has generally similar behavior  
	for models with $\mchi \ge 350\gev$.
	The only noticeable difference is in the absolute predictions for 
	detection rates for models with $\mchi > 500 \gev$, where instead of 
	${\Gamma^\mu}_{\rm IGC} < 10^{-4}\hgev$ one expects the rates to be 
	${\Gamma^\mu}_{\rm ACC} < 3\cdot 10^{-5}\hgev$.
	The highest predicted detection rates of $10^{-1}\hgev$ are 
	again correlated to a few models 
	which are unaffected  by coannihilation
\cite{Bednyakov:2002js}.  
	The SQC does not significantly contribute to reduction of the rates.

Figure~\ref{CS-f08} shows neutralino-proton 
	scattering cross sections for the scalar 
	(spin-independent) and axial (spin-dependent) interactions.
	As in the previous figures, the models with $\mchi \le 250\gev$ 
	are hardly affected by coannihilation, and 
	for the majority of those models both neutralino-proton and 
	neutralino-neutron scattering cross sections reach values  
	$\sigma_{\chi\,p } \le 10^{-17}\dtgev$ with the 
	maximal cross section of the order of $10^{-15}\dtgev$. 

	Cosmologically interesting models with 
	$\mchi \ge 250\gev$ were 
	influenced by coannihilations in the way discussed above, 
	and the maximal value of the neutralino-nucleon cross section 
	decreases from $10^{-15}\dtgev$ to $10^{-16}\dtgev$
	for the models with $\mchi > 500\gev$. 
	All in all 
	independently of neglecting or including of NCC, SQC and SLC,
	the maximal scalar scattering neutralino-nucleon 
	cross section was large as  
	$10^{-15}$--$10^{-14}\dtgev$.	

	The spin-dependent neutralino-nucleon cross sections 
	are typically higher than the spin-independent ones, 
	and we have found the maximal values 
$10^{-10}\dtgev$ for the axial neutralino-proton and 
$10^{-11}\dtgev$ for the axial neutralino-neutron scattering for the models 
	which are unaffected by the coannihilations. 
	The majority of cosmologically interesting models yield 
	axial neutralino-proton scattering cross sections in the range 
	$5\cdot 10^{-16}\dtgev < \sigma_{\chi\,p}  < 2\cdot 10^{-12}\dtgev$ 
	and  
	axial neutralino-neutron scattering cross sections in the range 
	$2\cdot 10^{-16}\dtgev < \sigma_{\chi\,n}  <8\cdot 10^{-13}\dtgev $
\cite{Bednyakov:2002js}. 
	The SQC contributes to reduction of the cross sections, 
	but not significantly again. 

	Due to the fact that capture of $\chi$  
	in the Sun (contrary to the Earth) 
	occurs also via spin-dependent $\chi\,p$ interaction,
	there are noticeable correlations between the highest  
	upgoing muon rates from $\chi\chi$ annihilation in the Sun
	and the highest values of 
	the axial neutralino-proton scattering cross sections
(Figs.~\ref{IND-f07}b) and \ref{CS-f08}b)).
	
Figure~\ref{RGe-f09} shows the expected direct detection 
	event rates calculated for a $^{73}$Ge detector
	when NCC, SQC, SLC, and ACC are taken into account. 
	For models with $\mchi \le 250\gev$ 
	coannihilations of any kind play no role.
	The estimations of the event rate for models with 
	$\mchi \ge 400\gev$ are decreased (to about 0.005 event/(kg day))
	due to NCC
\cite{Bednyakov:2002js}.

\section{Conclusion} 
	Due to continuing improvements of the accuracy of astrophysical data 
	and the importance of relic density constraints
	for SUSY models the precision calculation of 
	the neutralino relic density is very desirable.
	The progress in this direction is very fast.
	Recently a new sophisticated C code {\tt micrOMEGAs}
	on the basis of {\tt CompHEP}
\cite{comphep} for calculations of the relic density in the MSSM 
	has been presented
\cite{Belanger:2001fz}.
	It  includes all coannihilation channels with neutralinos, charginos, 
	sleptons, squarks and gluinos. 
	The relic density of neutralinos in the mSUGRA 
	was calculated on the basis of annihilation diagrams 
	involving sleptons, charginos, neutralinos and 
	third generation squarks in
\cite{Baer:2002fv}.
	This paper is mostly devoted to 
	prospects for SUSY search with 
	various $e^+e^-$ and hadron colliders 
	and pays no attention to the individual 
	contributions of different coannihilation channels.
	In addition, 
	a full set of exact, analytic expressions for the 
	annihilation of the lightest neutralino pairs 
\cite{Nihei:2002ij}
	as well as slepton-neutralino coannihilations
\cite{Nihei:2002sc}
	into all two-body
      	tree-level final states in the framework of minimal SUSY 
	is now available. 
	The authors of the DarkSusy have also made new efforts 
\cite{Baltz:2002ei} to include all possible 
	channels of coannihilations in their public-available 
	DarkSusy code.
	
	Following this main direction
	we calculated the neutralino relic density (RD)
 	taking into account slepton-neutralino (SLC), 
	neutralino-chargino/neutralino (NCC), and 
	squark-neutralino (SQC)	coannihilation channels
	within the low-energy effective MSSM. 
	To this end we have implemented in our code
\cite{BKKmodel} the
	relic density part (with neutralino-chargino coannihilations)
	of the DarkSusy code 
\cite{Darksusy} supplied with the adopted code of
\cite{EFOS-stau} (calculating slepton-neutralino coannihilations)
	and the generalized (including sbottom-neutralino coannihilations)
	code of 
\cite{Ellis:2001nx} (calculating stop-neutralino coannihilations).  
	With the help of this new 
	code, in contrast with previous considerations,
	we pay attention to the 
	interplay between different coannihilation channels in 
	the effMSSM.
\begin{figure}[h] 
\begin{picture}(100,80)
\put(-10,-75){\includegraphics{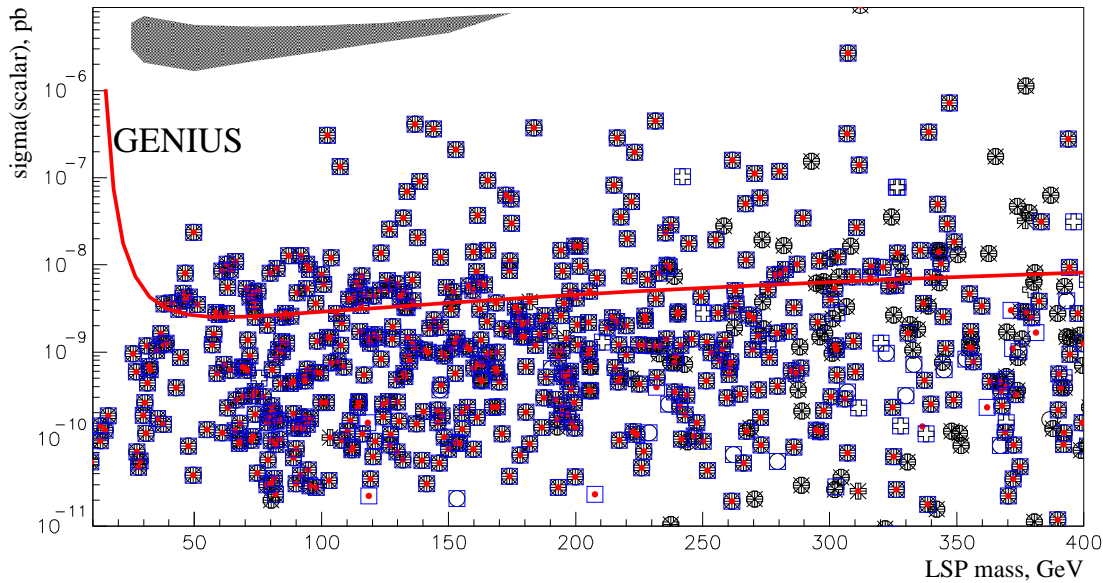}}
\end{picture}
\caption{Neutralino-proton scattering cross sections for 
	scalar (spin-independent) interaction.
	The notations are as in 
Fig.~\ref{IND-f07}.	
	Expectations for the GENIUS detector 
\protect\cite{GENIUS} and the annual-modulation region of DAMA 
	(shaded region) 
\protect\cite{DAMA} are also given.
	The maximal sensitivity of GENIUS and the region of DAMA are located 
	at $40 \le \mchi \le 300\gev$.
\label{GENIUS-f10}}
\end{figure} 

	We have shown that in the effMSSM the maximum factors of RD decrease 
	due to NCC as well as due to SQC and SLC can reach $10^{-3}$, 
	as long as the lower experimental limits for 
	${m_{\tilde \tau}}$, ${m_{\tilde t}}$, ${m_{\tilde b}}$, 
	and $m_{\tilde\chi^\pm}$, 
	are of the order of $80\gev$.
	We conclude that SQC, NCC, SLC produce comparable 
	RD reduction effects in the effMSSM.
	For the majority of models which are affected by coannihilations and  
	which successfully passed all relevant accelerator, 
	cosmological and rare-decay constraints
	it was observed that
	either NCC, SQC or SLC alone produces significant reduction of RD 
	while the other coannihilation channels give  
	considerably smaller or zero further reduction.
	Contrary to NCC and SLC, which 
	produce non-negligible effects only 
	if the NLSP mass is smaller than $1.15\mchi$, 
	for SQC the relevant NLSP mass could reach $1.50\mchi$. 
	The type of NLSP determines the dominant coannihilation channel
\cite{Bednyakov:2002js}. 
	Due to the fact that the effMSSM 
	more often favors neutralinos and charginos, but not 
	sfermions, to be the NLSP (the NLSP-LSP mass differences in general 
	are systematically larger for sfermions than for gauginos),
	the NCC channel more often dominantes in agreement with 
\cite{EdsjoGondolo}.
	Only LSPs with the gaugino fraction ${{\cal P}} > 0.7$ 
	remain CDM candidates of cosmological interest.
	
	Some models for which ${\Omega h^2 }>0.3$  
	when coannihilations is neglected fall within 
	the cosmologically interesting region merely due 
	to SLC and SQC, and some other models 
	shift out of the region below ${\Omega h^2 }<0.1$ merely due to NCC.
	In the effMSSM, contrary to mSUGRA 
\cite{EFOS-stau}, all coannihilations do not impose 
	new cosmological limits on the mass of the LSP
\cite{Bednyakov:2002js}.
	The optimistic predictions for neutralino-nucleon 
	cross sections, indirect and direct detection rates 
	for cosmologically interesting models 
	are almost unaffected by these coannihilations. 
	Only for large $\mchi \ge 400\gev$, 
 	the respectively high values are reduced, 
	mainly because the NCC excludes the corresponding 
	models from the cosmological interesting region 
	$0.1<\IGC<0.3$.

	Therefore, future increase of the 
	lower mass limits for all possible NLSPs 
	(at Tevatron or LHC) can, in principle, strongly 
	reduce the importance of the effect of any of the 
	coannihilation channels. 
	Furthermore, we would like to note that
	despite obvious importance of sophisticated RD
	calculations including complete sets of coannihilation channels, 
	it may happen that coannihilations play no role at least for 
	{\em direct}\ detection of cold dark matter. 
	From 
Fig.~\ref{GENIUS-f10}
	one can see that the field of 
	maximal sensitivity of the best new-generation  
	CDM detectors, like GENIUS
\cite{GENIUS}, as well as the annual-modulation region of DAMA 
\cite{DAMA}  are located 
	at $40 \le \mchi \le 300\gev$, where coannihilation
	effects are almost invisible.

\smallskip 
	The authors thank Yudi Santoso for making his code available
	and I.V. Krivosheina for permanent interest in the work. 
	V.B. thanks the Max Planck Institut fuer Kernphysik 
	for the hospitality and the RFBR 
	(Grants 00--02--17587 and 02--02--04009) for support.



\begin{thebibliography}{999}\vspace*{-3.5\baselineskip}
\bibitem{acc}
	{P. de Bernardis}, astro-ph/{0004404};
	{A. Balbi}, astro-ph/{0005124}.
\bibitem{kt90}
	E.W. Kolb and M.S. Turner, 
	{\em The Early Universe}, Addison-Wesley (1990);
	M.S. Turner, astro-ph/0108103.
\bibitem{jkg}  
	G.~Jungman, M.~Kamionkowski, and K.~Griest, 
	Phys.~Rep. {\bf 267} (1996) 195.
\bibitem{susyreview}
	H.P. Nilles, Phys.\ Rept. {\bf 110}, 1 (1984);
	H.E.~Haber and G.~Kane, Phys.\ Rept. {\bf 117}, 75 (1985);
	S. Martin, {hep-ph/9709356}. 
\bibitem{cmb} 
	A.~T.~Lee {\it et al.} (MAXIMA Collab.),
	Astrophys.\ J.\  {\bf 561}, L1 (2001); 
	C.B. Netterfield {\it et al.} (BOOMERANG Collab.), astro-ph/0104460; 
	N.W. Halverson {\it et al.} (DASI Collab.), astro-ph/0104489; 
	P. de Bernardis {\it et al.}, astro-ph/0105296,
	A.~Melchiorri, astro-ph/0201237.

\bibitem{goldberg83}
	H. Goldberg, Phys. Rev. Lett. {\bf 50}, 1419 (1983).
\bibitem{ehnos}
	J. Ellis, J. Hagelin, D. Nanopoulos, and M. Srednicki, 
	Phys. Lett. B {\bf 127}, 233 (1983);
	J. Ellis, J. Hagelin, D. Nanopoulos, K. Olive, and M. Srednicki, 
	Nucl. Phys. B {\bf 238}, 453 (1984).
\bibitem{krauss83}
	L.M.~Krauss, Nucl. Phys. B {\bf 227}, 556 (1983).
\bibitem{griest88}
	K. Griest, Phys. Rev. D {\bf 38}, {2357} ({1988});
	Erratum {\bf 39}, {3802} ({1989}).
\bibitem{gkt} 
	K. Griest, M. Kamionkowski, and M. Turner, 
	Phys. Rev. D {\bf 41}, 3565 (1990).
\bibitem{erl90}
	J. Ellis, L. Roszkowski, and Z. Lalak, 
	Phys. Lett. B {\bf 245}, {545} (1990).
\bibitem{os91}
	K.A. Olive and M. Srednicki, 
	Phys. Lett. B {\bf 230}, 78 (1989);
	Nucl. Phys. B {\bf 355}, 208 (1991).
\bibitem{DreesNojiri} 
	M. Drees and M. Nojiri, Phys. Rev. D {\bf 47}, 376 (1993).
\bibitem{GriestSeckel} 
	K. Griest and D. Seckel, Phys. Rev. D {\bf 43}, 3191 (1991).
\bibitem{GondoloGelmini} 
	P. Gondolo and G. Gelmini, Nucl. Phys. B {\bf 360}, 145 (1991).

\bibitem{an93} 
	P. Nath and R. Arnowitt, Phys. Rev. Lett. 
	{\bf 69}, 725  (1992); {\bf 70}, 3696 (1993);
	Phys. Lett. B {\bf 437}, 344 (1998).
\bibitem{lny93}
	J.L. Lopez, D.V. Nanopoulos, and K. Yuan, 
	Phys. Rev. D {\bf 48}, 2766 (1993).
\bibitem{ows} 
	M. Srednicki, R. Watkins, and K. Olive, 
	Nucl. Phys. B {\bf 310}, 693 (1988).
\bibitem{BaerBrhlik} 
	H. Baer and M. Brhlik, 
	Phys. Rev. D {\bf 53}, 597 (1996); {\bf 57}, 567 (1998); 
	H.~Baer \etal,  
	Phys.\ Rev.\ D {\bf 63}, 015007 (2001).
\bibitem{barb} 
	R. Barbieri, M. Frigeni, and G. F. Giudice, 
	Nucl. Phys. B {\bf 313}, 725 (1989).
\bibitem{Bottino} 
	A. Bottino, V. de Alfaro, N. Fornengo, G. Mignola, and S. Scopel, 
	Astropart. Phys. {\bf 1}, 61 (1992); 
	A. Bottino {\it et al.}, 
	Astropart. Phys. {\bf 2}, 67 (1994); 
	V. Berezinsky {\it et al.}, 
	Astropart. Phys. {\bf 5}, 1 (1996);
	A. Bottino, F. Donato, N. Fornengo, and S. Scopel, 
	Phys. Rev. D {\bf 59}, 095004 (1999).
\bibitem{leszek} 
	J. Ellis and L. Roszkowski, 
	Phys. Lett. B {\bf 283}, 252 (1992); 
	L. Roszkowski and R. Roberts, 
	Phys. Lett. B {\bf 309}, 329 (1993); 
	G. Kane, C. Kolda, L. Roszkowski, and J. Wells, 
	Phys. Rev. D {\bf 49}, 6173 (1994).

\bibitem{EdsjoGondolo} 
	J.~Edsjo and P.~Gondolo,
	Phys.\ Rev.\ D {\bf 56}, 1879 (1997);	
	Phys. Atom. Nucl. {\bf 61}, 1181, (1998);
	P.~Gondolo and J.~Edsjo,
	hep-ph/9804459;
	J.~Edsjo,
	hep-ph/9704384.
\bibitem{Darksusy} 
	P. Gondolo, J. Edsj\"o, L. Bergstr\"om, P. Ullio, and E.A. Baltz
	astro-ph/0012234; 
	http://www.physto.se/~edsjo/darksusy/. 

\bibitem{Ellis-Higgs} 
	J.~R.~Ellis, T.~Falk, G.~Ganis, K.~A.~Olive, and M.~Srednicki,
	Phys.\ Lett.\ B {\bf 510}, 236 (2001).
\bibitem{EFOS-stau} 
	J.~R.~Ellis, T.~Falk, and K.~A.~Olive,
	Phys.\ Lett.\ B {\bf 444}, 367 (1998);
	J.~R.~Ellis, T.~Falk, K.~A.~Olive, and M.~Srednicki,
	Astropart.\ Phys.\  {\bf 13}, 181 (2000);
	[Erratum-ibid.\  {\bf 15}, 413 (2000)].
\bibitem{Belanger:2001fz} 
	G.~Belanger, F.~Boudjema, A.~Pukhov, and A.~Semenov,
	hep-ph/0112278.

\bibitem{Gomez:2000sj}
	M.~E.~Gomez, G.~Lazarides, and C.~Pallis,
	Phys.\ Lett.\ B {\bf 487}, 313 (2000);
	Phys. Rev.  D {\bf 61}, 123512 (2000). 
\bibitem{Gomez:2000ck}
	M.~E.~Gomez and J.~D.~Vergados,
	Phys.\ Lett.\ B {\bf 512}, 252 (2001)
	[arXiv:hep-ph/0012020].

\bibitem{Lahanas:2000uy} 
	A.~B.~Lahanas, D.~V.~Nanopoulos, and V.~C.~Spanos,
	Phys.\ Rev.\ D {\bf 62}, 023515 (2000).
\bibitem{Arnowitt:2001yh} 
	R.~Arnowitt, B.~Dutta, and Y.~Santoso,
	Nucl.\ Phys.\ B {\bf 606}, 59 (2001); 
	{R. Arnowitt, B. Dutta}, hep-ph/{0112157}.
\bibitem{Corsetti:2001yq} 
	A.~Corsetti and P.~Nath,
	Phys.\ Rev.\ D {\bf 64}, 125010 (2001); 
	A.~Corsetti and P.~Nath,
	hep-ph/0005234.
\bibitem{Boehm:2000bj} 
	C.~Boehm, A.~Djouadi, and M.~Drees,
	Phys.\ Rev.\ D {\bf 62}, 035012 (2000).

\bibitem{Ellis:2001nx} 
	J.~R.~Ellis, K.~A.~Olive, and Y.~Santoso,
	hep-ph/0112113.

\bibitem{Belanger:2001am}
	G.~Belanger, F.~Boudjema, A.~Cottrant, R.~M.~Godbole, and A.~Semenov,
	Phys.\ Lett.\ B {\bf 519}, 93 (2001). 
\bibitem{Nihei:2001qs} 
	T.~Nihei, L.~Roszkowski, and R.~Ruiz de Austri,
	JHEP {\bf 0105}, 063 (2001). 
\bibitem{Baer:2002fv}
	H.~Baer, C.~Balazs, and A.~Belyaev,
	hep-ph/0202076.

\bibitem{Nihei:2002ij} 
	T.~Nihei, L.~Roszkowski and R.~Ruiz de Austri,
	JHEP {\bf 0203}, 031 (2002).

\bibitem{Mizuta:1993qp} 
	S.~Mizuta and M.~Yamaguchi,
	Phys.\ Lett.\ B {\bf 298}, 120 (1993). 


\bibitem{Nihei:2002sc}
	T.~Nihei, L.~Roszkowski and R.~Ruiz de Austri,
	JHEP {\bf 0207}, 024 (2002).

\bibitem{Santoso:2002xu}
	Y.~Santoso,
	arXiv:hep-ph/0205026.

\bibitem{Bednyakov:2002js}
	V.~A.~Bednyakov, H.~V.~Klapdor-Kleingrothaus and E.~Zaiti,
	Phys.\ Rev.\ D {\bf 66}, 015010 (2002).


\bibitem{BKKmodel} 
	V.~A.~Bednyakov and H.~V.~Klapdor-Kleingrothaus,
	Phys.\ Rev.\ D {\bf 63}, 095005 (2001);
	Phys.\ Rev.\ D {\bf 62}, 043524 (2000);
	V.~A.~Bednyakov, H.~V.~Klapdor-Kleingrothaus, and H.~Tu,
	Phys.\ Rev.\ D {\bf 64}, 075004 (2001).

\bibitem{Kim:2002cy}
	Y.~G.~Kim, T.~Nihei, L.~Roszkowski and R.~Ruiz de Austri,
	arXiv:hep-ph/0208069.

\bibitem{Ellis:2002rp}
	For current starus of CMSSM see
	J.~R.~Ellis, K.~Olive, and Y.~Santoso,
	hep-ph/0202110.
\bibitem{sugra} 
	A. Chamseddine, R. Arnowitt, and P. Nath,
	Phys. Rev. Lett. {\bf 49}, 970 (1982);
	R. Barbieri, S. Ferrara, and C. Savoy, 
	Phys. Lett. B {\bf 119}, 343 (1982);
	L.J. Hall, J. Lykken, and S. Weinberg, 
	Phys. Rev. {\bf D27}, 2359 (1983).
\bibitem{pdg} 
	K.~Hagiwara {\it et al.}  
	Phys.\ Rev.\ D {\bf 66}, 010001 (2002); 
	{http://pdg.web.cern.ch/pdg}.

\bibitem{flimits} 
	M. S. Alam \etal, (CLEO Collab.), 
	Phys.\ Rev.\ Lett.\ {\bf 74}, 2885 (1995); 
	K.~Abe {\it et al.},  (Belle Collab.), 
	hep-ex/0107065.
\bibitem{BerBorMasRi}
	S. Bertolini, F. Borzumati, A.Masiero, and G. Ridolfi, 
	Nucl. Phys. B {\bf 353}, 591 (1991); 
	R. Barbieri and G. Giudice, Phys. Lett. B {\bf 309}, 86 (1993);
	A. J. Buras \etal, Nucl. Phys. B {\bf 424}, 374 (1994);
	A. Ali and C.Greub, Z Phys. C {\bf 60}, 433 (1993).

\bibitem{progfalk} Toby Falk, private communication.
\bibitem{progsantoso} Yudi Santoso, private communication.

\bibitem{comphep} 
	A. Pukhov {\it et al.}, hep-ph/9908288;
	http://theory.sinp.msu.ru/\~{}pukhov/calchep.html.

\bibitem{Baltz:2002ei}
	E.~A.~Baltz and P.~Gondolo,
	arXiv:astro-ph/0207673.


\bibitem{GENIUS}
	H. V. Klapdor-Kleingrothaus, 
	{Int. Journal of Modern Physics} A {\bf 13}, 3953 (1998); 
	H. V. Klapdor-Kleingrothaus and Y. Ramachers. 
	{Eur. Phys. J.} A {\bf 3}, 85 (1998);
	H. V. Klapdor-Kleingrothaus et al., GENIUS: A Supersentive
	Germanium Detector System for Rare Events, Proposal, 
	{MPI-H-V26-1999}, August 1999, {hep-ph/9910205};
	H. V. Klapdor-Kleingrothaus, 
	{\it Springer Tracts in Modern Physics}, Vol. {\bf 163}, 69 (2000);
	H.~V.~Klapdor-Kleingrothaus,
	``GENIUS: A new underground observatory 
	for non-accelerator particle  physics,''
	arXiv:hep-ph/0206249.

\bibitem{DAMA} 
	R. Bernabei {\em et al.} (DAMA Collaboration), 
	{Phys. Lett.} B {\bf 480}, 23 (2000);
	{Eur. phys. J.} C {\bf 18}, 283 (2000).

\end{thebibliography}
\end{document}